# Surface-plasmon-polariton-driven narrow linewidth magneto-optics in Ni nanodisk arrays

Francisco Freire-Fernández,[1,*] Mikko Kataja,[1,2] and Sebastiaan van Dijken[1,*]

[1] *NanoSpin, Department of Applied Physics, Aalto University School of Science, P.O. Box 15100, FI-00076 Aalto, Finland*

[2] *Institut de Ciència de Materials de Barcelona (ICMAB-CSIC), Campus de la UAB, Bellaterra, Catalonia, Spain*

[*] E-mail: francisco.freirefernandez@aalto.fi; sebastiaan.van.dijken@aalto.fi

**Abstract:** Magnetoplasmonics exploits interactions between light and magnetic matter at the nanoscale for light manipulation and resonant magneto-optics. One of the great challenges of this field is overcoming optical losses in magnetic metals. Here we exploit surface plasmon polaritons (SPPs) excited at the interface of a $SiO_2$/Au bilayer to induce strong magneto-optical responses on the Ni nanodisks of a periodic array. Using a reference system made of Au nanodisks, we show that optical losses in Ni do hardly broaden the linewidth of SPP-driven magneto-optical signals. Loss mitigation is attained because the free electrons in the Ni nanodisks are driven into forced oscillations away from their plasmon resonance. By varying the $SiO_2$ layer thickness and lattice constant of the Ni nanodisk array, we demonstrate tailoring of intense magneto-optical Kerr effects with a spectral linewidth down to ~25 nm. Our results provide important hints on how to circumvent losses and enhance magneto-optical signals via the design of off-resonance magnetoplasmonic driving mechanisms.

# 1 Introduction

The integration of plasmonics and magneto-optics has led to the emergence of a new research field known as magnetoplasmonics [1-3]. The main goal of magnetoplasmonic is twofold. First, the use of magnetic materials in plasmonic structures enables active light manipulation at the nanoscale via field-controlled breaking of time-reversal symmetry. Examples include demonstrations of magnetic modulations of the surface plasmon polariton (SPP) wave vector in Au-Co-Au trilayers [4], magneto-optical transparency in magnetoplasmonic crystals [5], and magnetic-field-controlled routing of light emission from diluted-magnetic-semiconductor quantum wells [6]. Second, the excitation of surface plasmons in magnetic materials can be used to resonantly enhance and spectrally tailor their magneto-optical response [7-10]. This configurability is attractive for label-free biosensing [11] and ultrafast all-optical magnetic switching [12].

Despite its promise, magnetoplasmonics faces a challenge of overcoming optical losses [13-15]. This holds particularly true for nanostructures containing magnetic metals whose losses are significantly higher compared to noble metals. One of the loss mitigating strategies in noble-metal plasmonics involves near- or far-field coupling of nanostructures. In periodic two-dimensional nanoparticle arrays, for instance, radiative coupling between diffracted waves in the array plane and localized surface plasmon resonances (LSPRs) produces collective surface lattice resonances (SLRs) [16-19]. The full-width at half maximum (FWHM) of SLRs in Au or Ag nanoparticle arrays can be <10 nm [19], which is a significant improvement from 80 – 100 nm for LSPRs in isolated nanostructures of these noble metals. Based on these results, it seems logical to transfer the concept of collective resonances as a loss mitigating mechanism to magnetoplasmonics. Arrays of higher-loss magnetic nanoparticles have indeed been shown to support the excitation of SLRs, rendering resonant enhancements of the magneto-optical activity [10,20]. Despite this attractive property, the SLRs of magnetic lattices are typically one order of magnitude broader than those of noble metal systems.

Another loss reduction strategy in magnetoplasmonics involves the integration of noble and magnetic metals. Realizations such as near-field coupled vertical dimers [21-23] and far-field coupled checkerboard patterns [24] have shown a moderate narrowing of plasmon resonances compared to the all-magnetic structures. The loss properties of these hybrid magnetoplasmonic systems can be understood by considering the coupling of weakly and heavily damped oscillators that are driven near their resonance frequencies by an external force.

A few other options are available to circumvent losses in magnetoplasmonics. For instance, one could refrain from using magnetic metals altogether. Most notably, hybrid structures of noble metals and dielectric iron garnets combine narrow plasmonic resonances and strong magneto-optical responses [7,9]. Alternative loss compensation mechanisms involving magnetic metals that have been proposed recently include the integration of thin magnetic films with photonic crystals [25] and lasing in Ni nanodisk arrays overlaid with an organic gain medium [26].

In this work, we demonstrate a hybrid magnetic metal/noble metal magnetoplasmonic structure that avoids the inherently high optical losses of magnetic materials. Our system consists of a square Ni nanodisk array placed onto a $SiO_2$/Au bilayer. In this configuration, the array acts as grating coupler for the excitation of spectrally narrow SPPs at the $SiO_2$/Au interface. The slowly decaying near-field of the SPP mode on the dielectric side of the $SiO_2$/Au interface [27], in turn, drives the free electrons of the Ni nanodisks into forced oscillations away from their resonance frequency. Through electronic spin-orbit coupling in the perpendicularly magnetized Ni nanodisks, this produces an intense magneto-optical Kerr effect in an equally narrow spectral range.

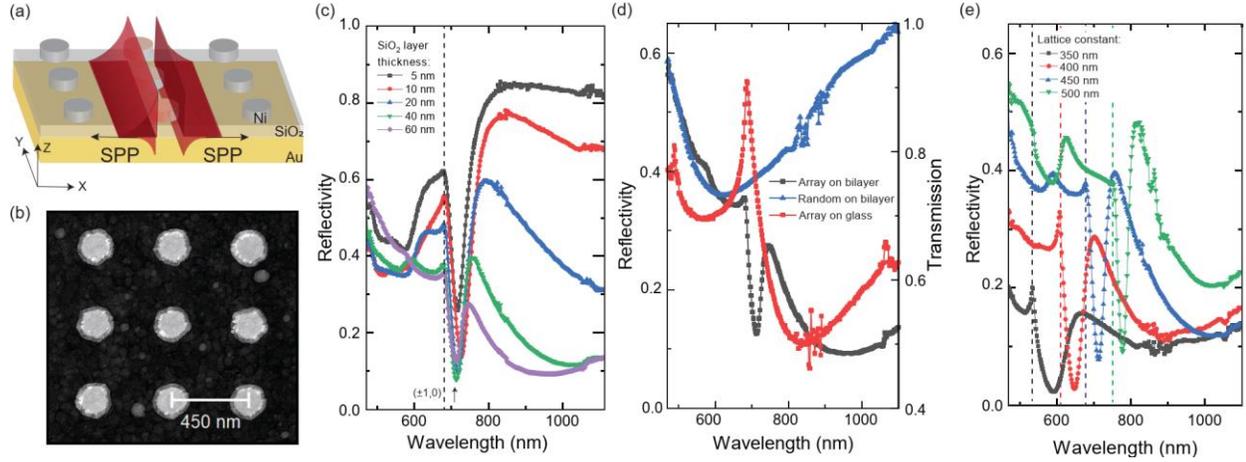

**Figure 1:** (a) Schematic of the magnetoplasmonic sample. (b) Atomic force microscopy (AFM) image of a Ni nanodisk array on top of a SiO$_2$/Au bilayer. (c) Optical reflectivity measurements of Ni nanodisk arrays on SiO$_2$/Au bilayers with $a$ = 450 nm and varying SiO$_2$ thickness. (d) Reflectivity spectrum of a Ni nanodisk array on top of a 60 nm SiO$_2$/Au bilayer and transmission curve of an identical Ni nanodisk array on glass. The lattice constant of the arrays is 450 nm. For comparison, the reflectivity spectrum of randomly distributed Ni nanodisks on a SiO$_2$/Au bilayer is shown also. (e) Optical reflectivity measurements of Ni nanodisk arrays on SiO$_2$/Au bilayers with a constant SiO$_2$ thickness of 40 nm and different lattice constants. The vertical dashed lines in (c) and (e) indicate the ($\pm 1$, 0) diffracted orders of the Ni nanodisk arrays.

## 2 Results and discussion

Figure 1(a) schematically illustrates our magnetoplasmonic structure. The 150-nm-thick Au film is grown by electron beam evaporation onto a glass substrate and it is covered by an atomic-layer-deposited SiO$_2$ film of varying thickness (5 - 60 nm). On top of this bilayer, the Ni nanodisk arrays are patterned using electron beam lithography. The Ni nanodisks have a diameter of 200 nm and a thickness of 70 nm and they are arranged into square arrays with a lattice constant ($a$) ranging from 350 nm to 500 nm (Figure 1(b)). As reference, we made identical all-noble-metal structures by replacing the Ni nanodisks with Au. We also fabricated Co/Pt nanodisk arrays on top of TiO$_2$/Au bilayers.

We start the discussion of results by considering optical reflectivity measurements. The samples are characterized at normal incidence using linearly polarized light along the *x*-axis of the nanodisk arrays (Figure 1(a)). To ensure a homogeneous dielectric environment, we embedded the nanodisks in index-matching oil (refractive index $n = 1.52$). Figure 1(c) shows the optical reflectivity of square Ni nanodisk arrays on top of $SiO_2$/Au bilayers. In these measurements, the lattice constant is fixed at 450 nm and the thickness of the $SiO_2$ is varied from 5 nm to 60 nm. The reflectivity spectra display maxima at the wavelength of the Rayleigh anomaly, $\lambda_{p,q} = \frac{na}{\sqrt{p^2+q^2}}$, corresponding to the $(p, q) = (\pm 1, 0)$ diffracted orders of the Ni nanodisk array, with *p* and *q* indicating the order of diffraction along *x* and *y*, respectively. If the wave vector of the diffracted waves in the array plane matches that of a plasmon mode at the $SiO_2$/Au interface, two counterpropagating SPPs are excited. Energy absorption by the SPP mode reduces the reflectivity. Because the wave vectors of the SPPs are given by $k_{SPP} = \pm k_0 \sqrt{\frac{\varepsilon_{SiO_2} \cdot \varepsilon_{Au}}{\varepsilon_{SiO_2} + \varepsilon_{Au}}}$, the free space wavelength of a SPP excitation corresponds to [28]

$$\lambda'_{p,q} = \frac{a}{\sqrt{p^2+q^2}} \cdot \sqrt{\frac{\varepsilon_{SiO_2} \cdot \varepsilon_{Au}}{\varepsilon_{SiO_2} + \varepsilon_{Au}}} . \qquad (1)$$

Here, $k_0$ is the wave vector in free space and $\varepsilon_{SiO_2}$ and $\varepsilon_{Au}$ are the dielectric constants of $SiO_2$ and Au. Based on Equation 1 and dielectric constants determined by ellipsometry, we estimate a SPP excitation wavelength of 706 nm at normal incidence. This value is congruent with our measurements indicating a deep minimum in the optical reflectivity spectra at a wavelength of ~710 nm. The resonances labeled by the upward-pointing arrow in Figure 1(c) thus correspond to the excitation of narrow-linewidth SPPs at the $SiO_2$/Au interface.

In addition to the SPP mode, the reflectivity spectra of Figure 1(c) exhibit a second resonance at larger wavelength. For a SiO$_2$ thickness of 60 nm, the spectral shape of this feature resembles that of an identical Ni nanodisk array on glass, as illustrated by the data of Figure 1(d). In the latter geometry, the resonance at 830 nm corresponds to the excitation of a SLR mode [10,20]. For the Ni nanodisk array on 60 nm SiO$_2$/Au, the resonance wavelength is shifted to 940 nm and it redshifts further for thinner SiO$_2$ layers (see Figure 1(c)). Simultaneously, the resonance mode loses intensity. Both effects are explained by the formation of image dipoles in the Au film if the SiO$_2$ layer is thin, as discussed previously for noble-metal plasmonic systems [29-34]. The collective nature of the plasmon resonance mode in the Ni nanodisk arrays on top of SiO$_2$/Au bilayers is confirmed further by the distinct optical response of randomly distributed Ni nanodisks on SiO$_2$/Au (blue curve in Figure 1(d)). In the latter case, the broad resonance centered at ~620 nm corresponds to the LSPR mode of individual Ni nanodisks. Once placed in a periodic array, the LSPRs couple and the resonance wavelength redshifts. Hereafter, we refer to this collective resonance as the SLR-related mode.

Figure 1(e) summarizes the dependence of the optical reflectivity spectra on lattice constant for a fixed SiO$_2$ thickness of 40 nm. As anticipated, the SPP and SLR-related modes shift to larger wavelength when the (±1, 0) diffracted order moves up. The reflectivity minima closely follow the SPP excitation wavelength given by Equation 1 for all samples. An increase of the lattice constant from 350 nm to 500 nm narrows the SPP resonance. For $a = 500$ nm, the SPP mode in the reflectivity spectrum has a FWHM of 25 nm.

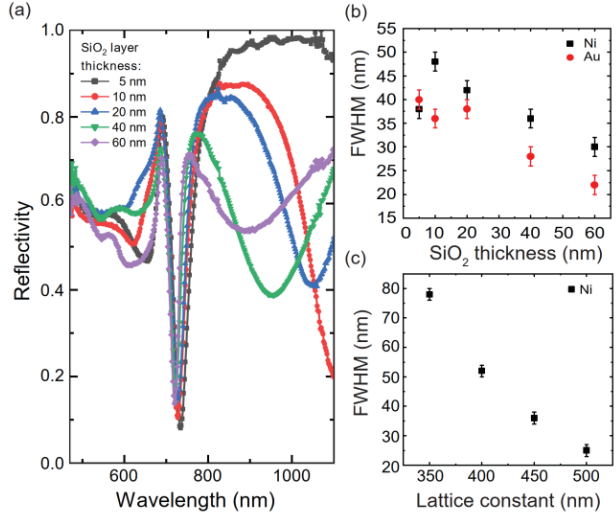

**Figure 2:** (a) Optical reflectivity measurements of Au nanodisk arrays on $SiO_2$/Au bilayers with $a = 450$ nm and varying $SiO_2$ thickness. (b) FWHM of the SPP mode in reflectivity spectra of Ni and Au nanodisk arrays with $a = 450$ nm on $SiO_2$/Au bilayers with varying $SiO_2$ thickness. (c) Dependence of the SPP resonance linewidth on lattice constant for Ni nanodisk arrays on $SiO_2$/Au. The $SiO_2$ layer thickness is 40 nm.

We now address the optical losses in our hybrid magnetoplasmonic structure. As already pointed out, the introduction of magnetic metals into noble-metal plasmonic systems tends to broaden the resonances because of mode hybridization [21-24]. Coupling between the narrow-linewidth SPP mode and the broader plasmon resonance on the Ni nanodisks could have a similar effect. To investigate the potentially detrimental role of optical losses in Ni, we fabricated identical samples with Au nanodisks. Figure 2(a) shows the optical reflectivity of the reference structures with $a = 450$ nm and different $SiO_2$ thickness. The spectra resemble those of the samples with Ni nanodisks depicted in Figure 1(c). Particularly, the FWHM of the SPP resonances in the reflectivity spectra are similar for both plasmonic systems (Figure 2(b)). This weak dependence on material demonstrates that the SPP and localized mode do not significantly hybridize. Optical losses in the Ni nanodisks therefore do not affect the SPPs much. Figure 2(b) summarizes the FWHM of the SPP resonance as a function $SiO_2$ thickness in both systems and Figure 2(c) depicts its dependence on lattice constant for the Ni nanodisk samples. We note that the minor effect of optical losses in the Ni nanodisks on the SPP linewidth is universal. To illustrate this further, we fabricated arrays of Co/Pt nanodisks on top of $TiO_2$/Au bilayers. The results shown in Figure S1 of the Supplementary Material indicate that SPP resonances with a FWHM down to ~25 nm are attained for $a$ = 450 nm and $t_{TiO_2}$ = 30 nm, which compares well to the data for the Ni and Au nanodisk samples (Figure 2(b)).

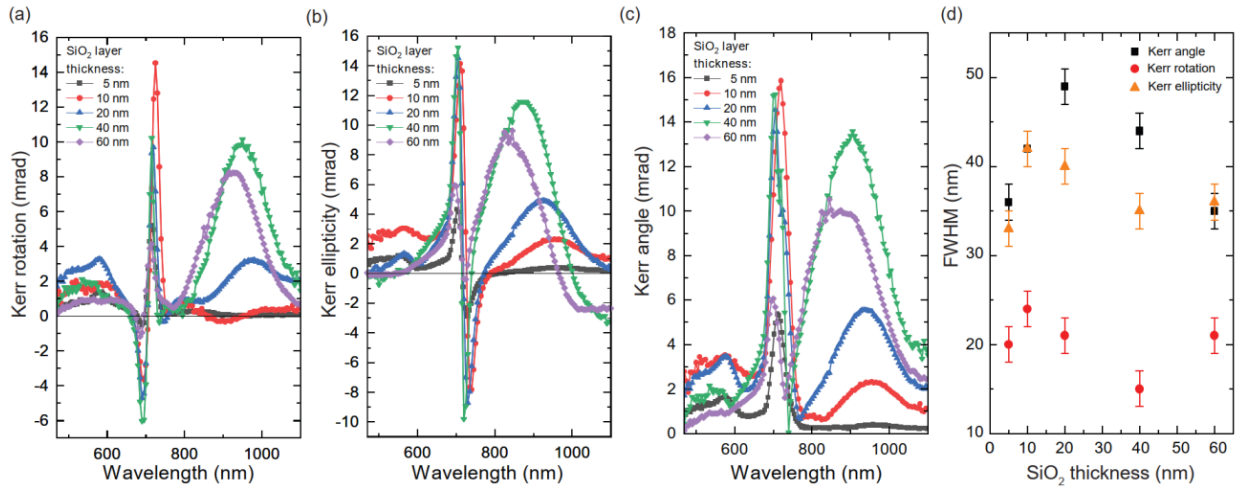

**Figure 3:** (a-c) Magneto-optical Kerr rotation, Kerr ellipticity, and Kerr angle of Ni nanodisk arrays on $SiO_2$/Au bilayers with $a$ = 450 nm and varying $SiO_2$ thickness. (d) FWHM of the SPP-driven resonance in the magneto-optical spectra of (a-c).

Figure 3 depicts how the excitation of plasmon resonances translates to the magneto-optical activity of the Ni nanodisk samples. In (a) and (b), we plot the polar Kerr rotation ($\theta$) and Kerr ellipticity ($\varepsilon$) for square nanodisk arrays with $a$ = 450 nm and different $SiO_2$ thickness, i.e., the same samples as in Figure 1(c). Both parameters are measured simultaneously using a photoelastic modulator and lock-in detection (see Methods). Additionally, we plot the Kerr angle $\Phi = \sqrt{\theta^2 + \varepsilon^2}$ in Figure 3(c). The measurements reveal that the SPPs at the $SiO_2$/Au interface produce an intense and narrow-linewidth magneto-optical signal. At the SPP wavelength of ~710 nm, the Kerr rotation peaks and the Kerr ellipticity changes sign.

To rationalize the magneto-optical response, one has to consider how the optical near-field of the leaky SPP mode drives free electrons in the Ni nanodisks into forced oscillations. The forced oscillations produce an electric dipole along the *x*-axis ($p_x$). Because we saturate the Ni magnetization perpendicular to the sample plane by an external magnetic field, electronic spin-orbit coupling within the Ni nanodisks produces a second electric dipole along the *y*-axis ($p_y$) [8]. The real and imaginary parts of the ratio between scattered fields along the *y*- and *x*-axis determine the Kerr rotation and Kerr ellipticity, respectively. Because of the phase dependence of these forced oscillations, the Kerr rotation and Kerr ellipticity signals in Figure 3(a,b) are spectrally more narrow than the SPP resonances in the reflectivity measurements of Figure 1(c) (this is particularly true for the Kerr rotation). In contrast, the FWHM of the SPP-driven resonances in the Kerr angle spectra are comparable to the reflectivity data. Figure 3(d) summarizes the linewidth of the SPP-driven magneto-optical signals.

The magnitude of the Kerr angle at the SPP wavelength is largest for 10-40 nm $SiO_2$ and drops when the Ni nanodisks are placed 60 nm above the $SiO_2$/Au interface (Figure 3(c)). The evanescent confinement of the SPP mode explains this fall off [27]. We also note a reduction of the Kerr signal for 5 nm $SiO_2$, which correlates with an increase in optical reflectivity (Figure 1(c)). This points to a weakening of the SPP-driven electric dipole on the Ni nanodisks when the $SiO_2$ layer is very thin. Screening of the electric dipole by the Au film, which is most efficient at small $SiO_2$ thickness, explains this effect. Besides the SPP resonance, the broader SLR-related mode also tailors the magneto-optical response (Figure 3(c)). Since the electric dipoles on the Ni nanodisks are primarily driven by the incident light at larger wavelengths, the dependence of the Kerr angle on $SiO_2$ thickness is different for this mode. The formation of image dipoles in the Au film almost completely suppresses the magneto-optical response for 5 nm $SiO_2$. Beyond this, the Kerr signal first increases to a maximum at $t_{SiO_2} = 40$ nm before declining again. The latter observation indicates that placing Ni nanodisk arrays onto an optimized dielectric/metal bilayer enhances their magneto-optical signal compared to an identical array on glass. We discuss this comparison in more detail later. Magneto-optical data for samples with varying lattice constant corroborate the main findings (Figure S2 and S3, Supplementary Material).

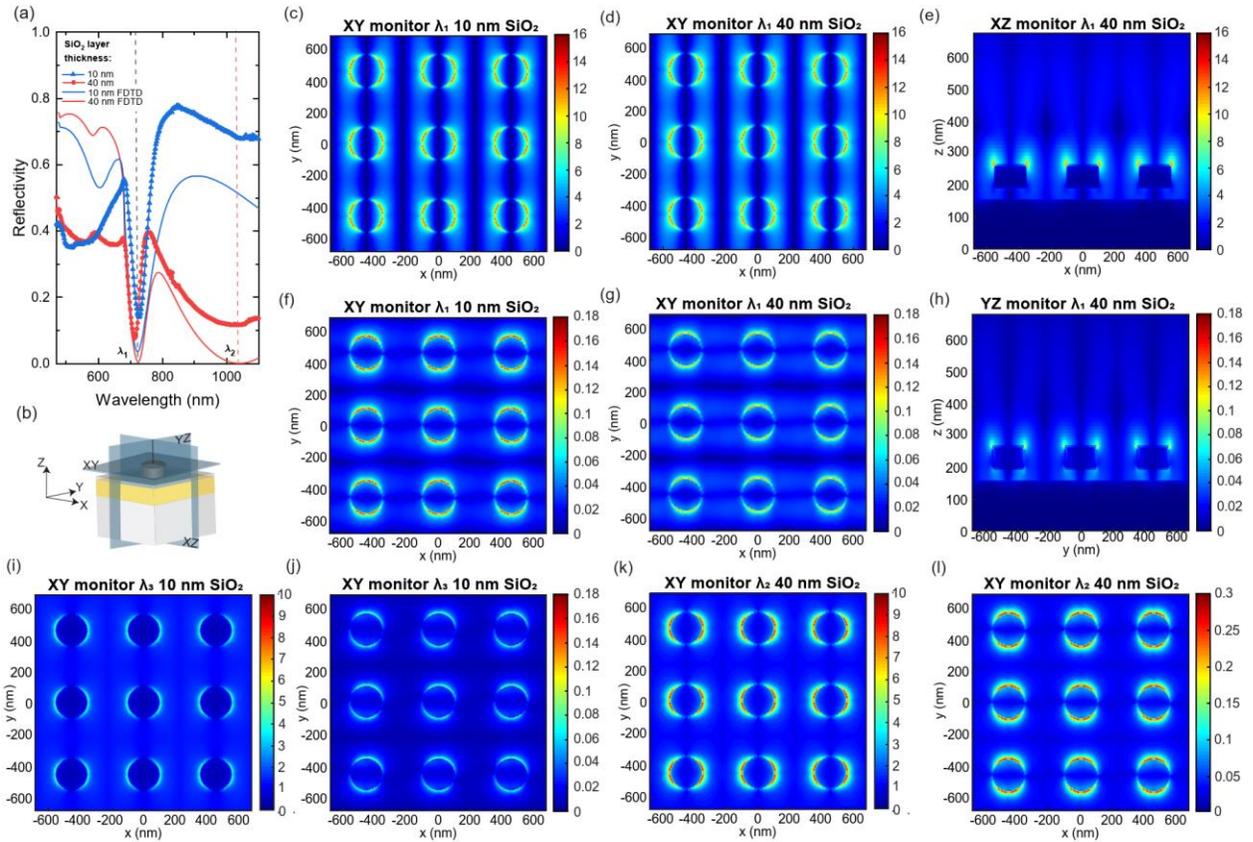

**Figure 4:** FDTD simulations of the magnetoplasmonic samples. (a) Comparison of simulated and measured reflectivity spectra of Ni nanodisk arrays on $SiO_2$/Au bilayers with $a = 450$ nm and a $SiO_2$ thickness of 10 nm and 40 nm. (b) Schematic of the monitors used to visualize the distribution of electric fields in different planes of the sample. (c-e) Electric field distribution at the SPP wavelength ($\lambda_1$) for samples with 10 nm $SiO_2$ and 40 nm $SiO_2$. (f-h) Field distribution of the orthogonal electric dipoles at $\lambda_1$ that are induced by spin-orbit coupling within the Ni nanodisks. (i-

l) Electric field distribution of the two orthogonal dipoles at the SLR-related resonance wavelength for samples with 10 nm SiO$_2$ ($\lambda_3$ = 1385 nm) and 40 nm SiO$_2$ ($\lambda_2$ = 1100 nm).

Finite-difference time-domain (FDTD) simulations in Lumerical software support our experimental results. In the simulations, we irradiated a 3 × 3 nanodisk array on top of a SiO$_2$/Au bilayer under periodic boundary conditions. This structure was excited at normal incidence by a plane wave with linear polarization along the *x*-axis (see Methods). Figure 4(a) compares reflectivity spectra for Ni nanodisk arrays on SiO$_2$/Au bilayers with $a$ = 450 nm and 10 nm or 40 nm of SiO$_2$. The simulated curves reproduce the main features of the experimental resonances. To visualize the distribution of electric fields at the resonance wavelengths, we placed a *xy* monitor right on top of the Ni nanodisks and *xz* and *yz* monitors intersecting the disks (Figure 4(b)). The monitors record two field contributions at the SPP wavelength ($\lambda_1$), the propagating SPPs and a localized mode on the Ni nanodisks. The SPP-driven electric dipoles on the Ni nanodisks have similar intensities for 10 nm SiO$_2$ and 40 nm SiO$_2$ (Figure 4(c-e)) and, through spin-orbit coupling, this also produces similar electric dipoles along the *y*-axis of the disks (Figure 4(f-h)). The simulated electric field strengths explain the large magneto-optical Kerr effect at the SPP wavelength of the corresponding experimental samples (Figure 3(a-c)). At the SLR-related resonance wavelength (Figure 4(i-l)), in contrast, the simulated electric fields are stronger for 40 nm SiO$_2$ than 10 nm SiO$_2$, in agreement with the experimental Kerr effect data. FDTD simulations for Ni nanodisk arrays on SiO$_2$/Au bilayers with a SiO$_2$ thickness of 5 nm are shown in Figure S4 of the Supplementary Material.

The effect of optical losses in the Ni nanodisks on the resonance properties of the SPP-driven electric dipole can be understood qualitatively by considering two coupled oscillators (Figure S5, Supplementary Material). One of the oscillators represents the surface plasmon mode on the Ni nanodisks. Coupling between single-particle LSPRs and the diffracted order of the periodic array and the formation of image dipoles in the Au layer determine the resonance properties of this mode. In the reflectivity spectra, this produces the SLR-related resonance at large wavelength. The second oscillator represents the SPP mode that is excited at the $SiO_2$/Au interface. If the resonance frequencies of the two oscillators are different, the SPP oscillator induces a narrow resonance on the Ni nanodisk oscillator. Importantly, the linewidth of this forced oscillation does not depend on the damping parameter of the driven oscillator (Figure S6 and S7, Supplementary Material). This condition is met in our hybrid magnetoplasmonic structures because of the clear spectral separation between the SPP and SLR-related resonances. In other words, off-resonance driving of the Ni nanodisk array by the near-field of a spectrally narrow SPP mode is hardly affected by the optical losses in Ni. For other hybrid magnetoplasmonic systems comprising noble and magnetic metals, the situation is often different. We illustrate this by considering vertical dimers and checkerboard arrays. In vertical dimers, two metal nanodisks of similar size couple via optical near-fields. The LSPRs of magnetic nanodisks are particularly broad making a spectral overlap of the two resonance modes in magnetoplasmonic dimers inevitable. In checkerboard arrays, lower and higher loss metal nanodisks couple via optical far-fields. In both geometries, the noble metal and magnetic subsystems can be represented by two coupled oscillators that are driven on-resonance (or close to resonance) by an external field. This leads to mode hybridization and linewidth broadening, the extend of which depends on the damping parameter of both oscillators (Figure S7, Supplementary Material).

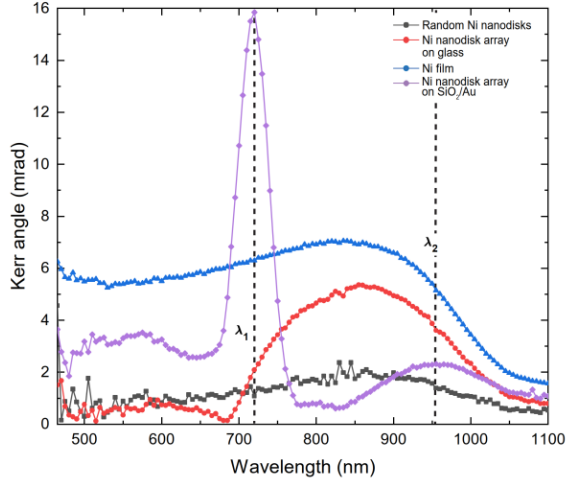

**Figure 5:** Magneto-optical Kerr angle measured on a random distribution of Ni nanodisks (black squares), a continuous 70 nm thick Ni film (blue triangles), a Ni nanodisk array on glass (red dots), and a Ni nanodisk array on a SiO$_2$/Au bilayer (purple diamonds). The Ni nanodisks are 70 nm thick and their diameter is 200 nm. The lattice constant of the arrays is 450 nm. The SiO$_2$ thickness is 10 nm.

To put our results into further perspective, we conclude this work by comparing different Ni nanodisk structures with identical shape, size, and packing density (Figure 5). For a random distribution of Ni nanodisks, the broad LSPR produces a weak and broad Kerr signal [8]. Ordering of the Ni nanodisks into a periodic array on glass enhances the Kerr angle through the excitation of a collective SLR [10,20]. The FWHM of the SLR mode is, however, broad (~200 nm). Placing the same Ni nanodisk array on a SiO$_2$/Au bilayer generates two main resonance in the magneto-optical spectrum, one broad SLR-related mode and a narrow SPP-driven mode. The intensities of both resonances can be tailored to about three times the signal of the Ni nanodisk array on glass by variation of the SiO$_2$ thickness (Figure 3). The narrow linewidth of the SPP-driven mode is particularly interesting for label-free magnetoplasmonic biosensing [11,35,36]. As outlined for randomly distributed Ni nanodisks in Ref. 11, phase-sensitive detection of the Kerr ellipticity nulling condition already provides a large enhancement of the refractometric sensing figure-of-merit. The sharp spectral features in the magneto-optical spectra of the samples presented here, would further increase the sensing performance by an estimated one to two orders of magnitude. Moreover, efficient time-reversal symmetry breaking in Ni nanodisk arrays on SiO$_2$/Au may be used to lift the degeneracy of plasmon modes at high-symmetry points of the Brillouin zone [37], which could lead to topological photonic effects.

# 3 Conclusions

In summary, we demonstrated narrow-linewidth magneto-optical Kerr signals by placing Ni nanodisk arrays within the optical near-field of SPPs at the interface of a SiO$_2$/Au bilayer. The low-loss magneto-optical resonances are explained by SPP-driven forced oscillations of free electrons and spin-orbit coupling in the Ni nanodisks. The wavelength and magnitude of the Kerr angle can be tailored by variation of the lattice constant and thickness of the SiO$_2$ layer.

# 4 Methods

**Sample fabrication**

The samples were fabricated on glass substrates using electron beam evaporation for the 150-nm-thick Au film and atomic layer deposition (ALD) for the SiO$_2$ layer. ALD was performed at 120 °C in a Beneq FTS 500 plasma system. The dielectric constants of the Au and SiO$_2$ films were measured using a J.A. Woollam ellipsometer. On top of the SiO$_2$/Au bilayers, the Ni nanodisk arrays were patterned using an electron beam lithography process in a Vistec EPBG5000pES system. After patterning of the PMMA resist layer, 70 nm of Ni was grown by electron beam evaporation and lift-off in acetone. For the reference samples, we used the same process with Au. We also fabricated Co/Pt nanodisks arrays on top of TiO$_2$/Au bilayers.

**Optical and Magneto-Optical Measurements**

For sample characterization, we used a magneto-optical Kerr spectrometer. The setup consisted of a NKT SuperK EXW-12 supercontinuum laser with an acousto-optical filter, polarizing and focusing optics, a Hinds Instruments I/FS50 photoelastic modulator, and a photodetector. The wavelength of the laser was tuned between 475 nm and 1100 nm. We used linear polarized light along the *x*-axis of the sample at normal incidence. During measurements, a ±400 mT field from an electromagnet switched the magnetization of the Ni nanodisks between the two perpendicular directions. The magneto-optical Kerr rotation and ellipticity were simultaneously recorded by lock-in amplification of the modulated signal at 50 kHz and 100 kHz. The optical reflectivity was determined using $R = 1 + \frac{I_{ref}-I}{I_{ref}-I_d}$, where $I$ and $I_{ref}$ are the reflected light intensities from the nanodisk array and an area without nanodisks, respectively, and $I_d$ is the dark current of the photodetector. All measurements were performed with the Ni, Au, or Co/Pt nanodisks immersed in index-matching oil with $n = 1.52$.

**FDTD Simulations**

Commercial Lumerical FDTD software was used to simulate the optical response of the magnetoplasmonic system. Magneto-optical effects were included in the simulations by following the procedure given in Ref. 38. For the material parameters of Ni and Au we used the data of Refs. 39 and 40, respectively. Only the Ni nanodisks were assumed to be magneto-optically active and contain off-diagonal permittivity tensor elements. We simulated 3 × 3 nanodisks on top of a SiO$_2$/Au bilayer and used periodic boundary conditions at the edges of the simulation area to mimic an extended array. The diameter and thickness of the Ni nanodisks and the thicknesses of the SiO$_2$ and Au layers were set to the experimental values. Moreover, the nanodisks were assumed to be surrounded by an index matching environment ($n = 1.52$). The sample was excited by a normal incident plane wave with linear polarization along the *x*-axis. The orthogonal electric dipoles depicted in Figure 4(f-h,j,l) were obtained by subtracting the simulation data for two perpendicular magnetization directions, which was implemented by exchanging the signs of the off-diagonal terms in the Ni dielectric tensor. Excitation of these orthogonal dipoles via spin-orbit coupling inside the Ni nanodisks produces the magneto-optical Kerr effect.

**Acknowledgments:** This work was supported by the Academy of Finland under Project No. 316857 and by the Aalto Centre for Quantum Engineering. Lithography was performed at the Micronova Nanofabrication Centre, supported by Aalto University.